\documentclass[a4paper,11pt]{article}
\usepackage{pos}

\newlength{\bibitemsep}
\newlength{\bibparskip}
\let\oldthebibliography\thebibliography
\renewcommand\thebibliography[1]{%
  \oldthebibliography{#1}%
  \setlength{\parskip}{\bibitemsep}%
  \setlength{\itemsep}{\bibparskip}%
}

\title{HIP and HEP}

\author{Urs Achim Wiedemann}

\affiliation{CERN Department of Theoretical Physics, CH-1211 Geneva}

\emailAdd{Urs.Wiedemann@cern.ch}

\abstract{This write-up of the ICHEP plenary ``Heavy Ions - theory''  focusses on some recent LHC discoveries and future opportunities
in heavy ion physics (HIP) that are at the intersection with high energy physics (HEP).  }

\FullConference{%
  40th International Conference on High Energy physics - ICHEP2020\\
  July 28 - August 6, 2020\\
  Prague, Czech Republic (virtual meeting)
}

\begin{document}
\maketitle

High energy physics (HEP) investigates the elementary constituents of matter and their interactions, but part
of its motivation and guidance arises from embedding HEP in the thermal history of our Universe.
For instance, the problem of baryon asymmetry motivates searches for baryon number violating processes and additional 
C- and CP-violating contributions in the fundamental interactions, but its solution needs to satisfy  all three Sakharov conditions.
Understanding the thermal equilibrium
and --  in this case -- out-of-equilibrium dynamics of fundamental quantum fields is an integral part of solving this problem, 
Similar comments apply to the further investigation of the dynamics 
of spontaneous symmetry breaking and the origin of mass. Embedding HEP in a thermal scenario is also an important element 
of HEP's hypothesis-building, as illustrated by the prevalent 
arguments about thermal relic abundances that inspire dark matter scenarios. In this wider sense, HEP 
investigates not only the elementary interactions, but it also touches upon the question: 
{\it How do non-abelian quantum field theories give rise to fundamental equilibrium and out-of equilibrium properties? How do
macroscopic, collective phenomena  arise from the fundamental interactions?}

For the strong interactions, these questions are in the focus of heavy ion physics (HIP). HIP grew out of HEP 
shortly after the formulation of quantum chromodynamics when theorists including Cabbibo and Parisi~\cite{Cabibbo:1975ig}, 
Collins and Perry~\cite{Collins:1974ky}, and Shuryak~\cite{Shuryak:1978ij} realized that Hagedorn's limiting temperature of a hadronic
system marks the transition temperature to the quark gluon plasma (QGP) phase. Today, the commonalities of 
HEP and HIP extend well beyond this intellectual starting point.  {\it Sociologically}, HIP and HEP
are united in the same collaborations at the LHC and they investigate the same data sets with an emphasis on arguably different questions but with common standards of quality and validation. {\it  Technically}, HIP and HEP work with the same accelerator chain, they share responsabilities for the same detectors, and they have common R{\&}D projects. The somewhat shorter planning periods of heavy ion projects makes HIP well-positioned to spearhead novel detector technologies~\cite{Adamova:2019vkf} and to explore novel ideas of using the versatile LHC machine~\cite{Aidala:2019pit}. {\it Strategically}, HIP has contributed to the HEP-led  European Particle Physics Strategy Update~\cite{Citron:2018lsq,Benedikt:2018csr}, and it is included for the first time in the Snowmass 2021 process that aims at documenting a vision for the future of particle physics in the U.S.~\cite{snowmass2021};  it may contribute beyond HL-LHC with $O(1000)$ collaborators to a future facility. {\it Scientifically},  the first decade of LHC has highlighted repeatedly topics at the interface between traditional HEP and traditional HIP physics. This plenary discusses some of them. 
%
\begin{figure}[t]
  \centering
  \vspace{-.5cm}
\includegraphics[width=0.49\linewidth]{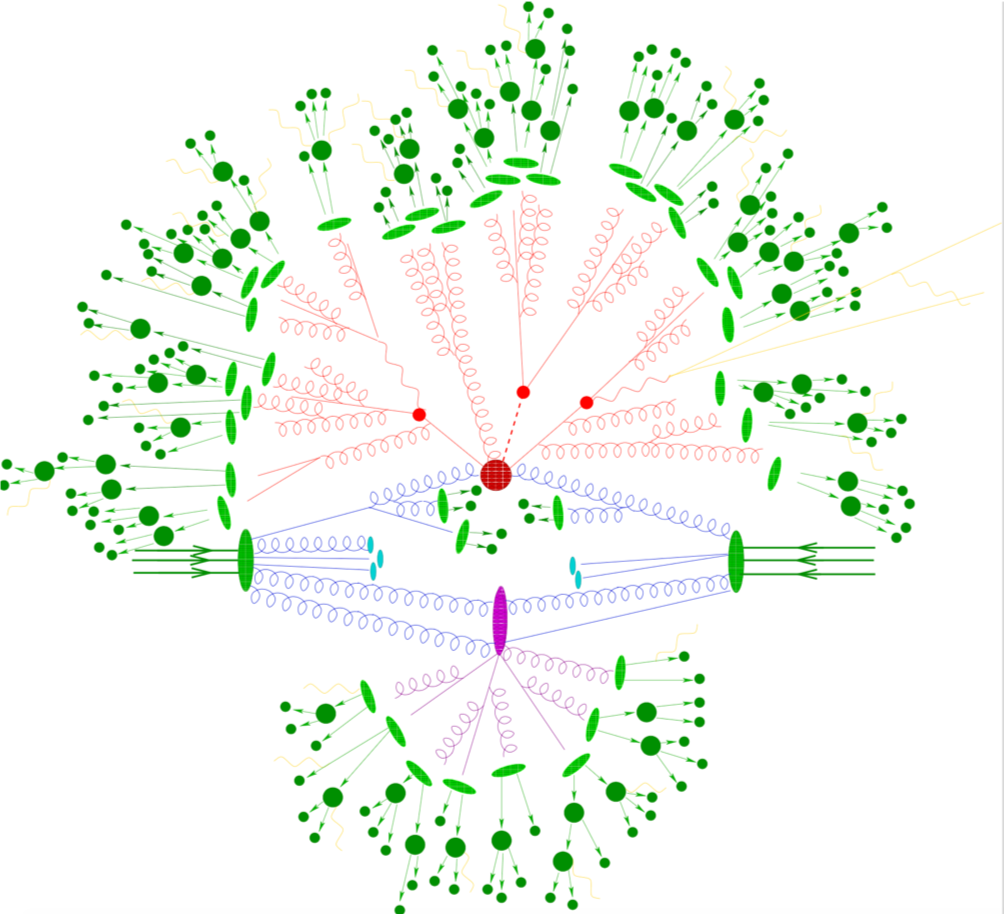}%
  \includegraphics[width=0.455\linewidth]{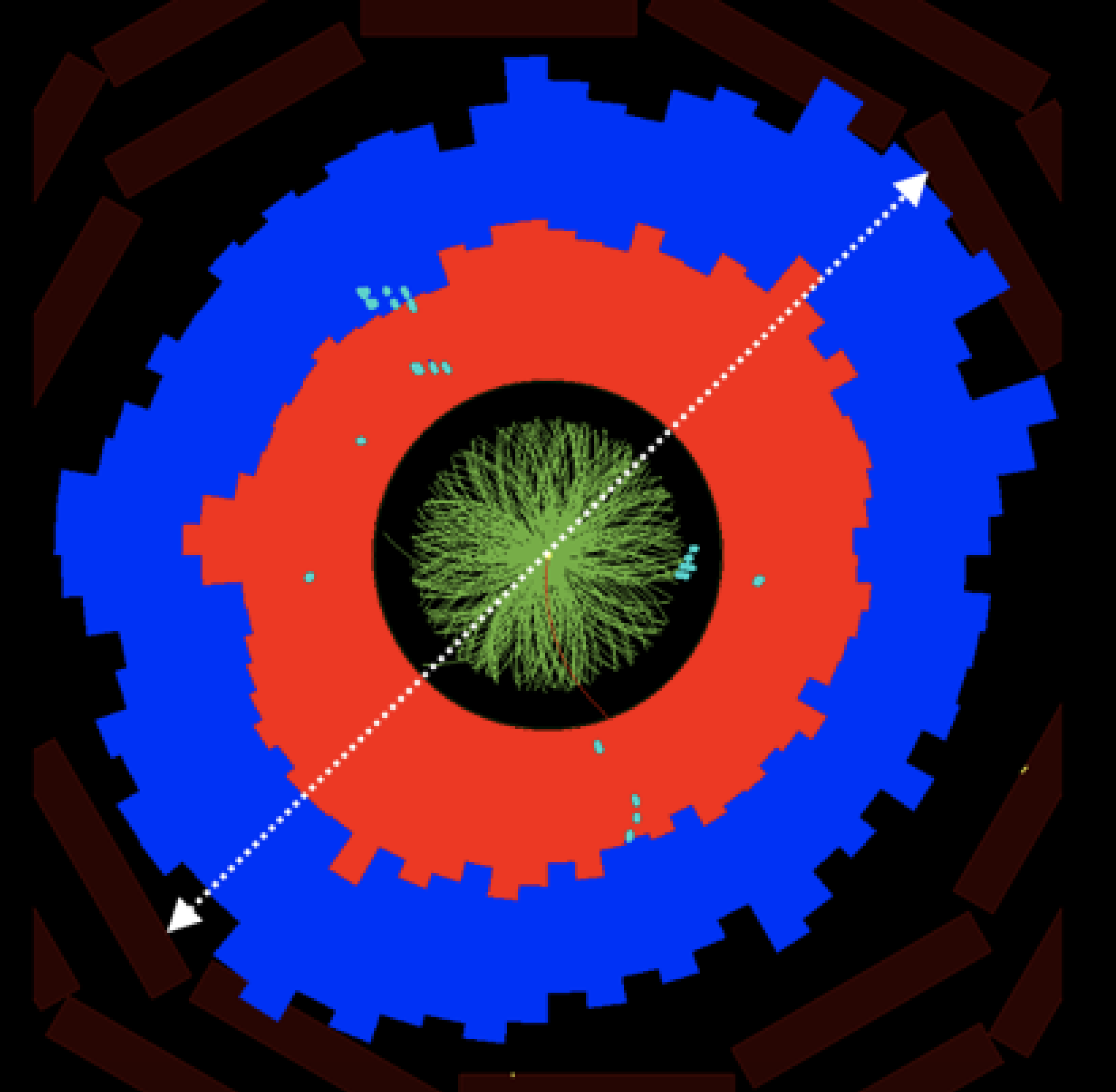}%
  \caption{Left: Schematic view of the physics implemented in standard multi-purpose event generators 
  for proton-proton collisions.
  Right: CMS event display of the calorimeter deposition of 
  more than 10'000 particles produced in a single, sufficiently central 2.76 TeV PbPb collision (plot from G. Roland, private communication). 
  The physics mechanisms sketched on the left cannot account for the azimuthal asymmetry of the calorimetric distribution displayed on the right.
}
  \label{fig1}
\end{figure}

The visible matter content of our Early Universe witnessed the most dramatic reduction of its effective physical degrees 
of freedom at the QGP transition at a temperature  $T_c \sim 2\, \times 10^{12}$ K at $O(10)$ microseconds after the Big Bang. 
Unfortunately, cosmological QGP signatures have remained elusive so far, though it is not excluded that they will become accessible 
in the upcoming era of precision cosmology~\cite{Caprini:2010xv,Boeckel:2011yj}. In hadronic collision experiments, however, the
energy densities required for QGP formation can be reached. 
By now, we have mature theoretical predictions for the temperature $T_c$ of the QGP transition, the steepness of its cross-over and many 
thermodynamic observables~\cite{Bazavov:2011nk,Borsanyi:2010bp,Ghiglieri:2018dib}. 
The much higher precision of LHC data (full feed-down corrections of hadrochemical abundances, precise data on higher harmonics $v_n$ and their kinematic dependencies, etc., see the ICHEP plenary~\cite{Pachmayer:2020apr}) motivates refined comparisons with finite temperature (``hot'') QCD. 
For {\it all} phase-space integrated hadrochemical ratios, a two-parameter model of statistical
hadronization arrives with small uncertainties at a temperature that is consistent with the transition temperature predicted 
by lattice QCD~\cite{Andronic:2017pug}. Global Bayesian analyses of larger and more differential data sets constrain hydrodynamic model parameters~\cite{Bernhard:2019bmu,Devetak:2019lsk,Auvinen:2020mpc,Nijs:2020roc} that have been
related to fundamental thermal properties such as $T_c$ and shear ($\eta/s$) and bulk ($\xi/s$) viscous transport 
coefficients. The methodology with which these quantities are related to data shares close similarities
with the analysis of the $\Lambda$CDM cosmological concordance model that features in the Particle Data Book. In HIP (in cosmology), 
the propagation of matter perturbations is used to constrain material properties of the smallest (the largest) material physical system 
studied by mankind. Given that $T_c$, $\eta/s$, $\xi/s$ are {\it fundamental} properties that can be calculated from first principles 
of thermal QCD, and given that current phenomenological analyses~\cite{Andronic:2017pug,Bernhard:2019bmu,Devetak:2019lsk,Auvinen:2020mpc,Nijs:2020roc,Shen:2020gef,Gajdosova:2020nvb}  
seem mature, the question arises whether the values derived phenomenologically from LHC and RHIC data could be included in the 
compilation of the Particle Data Group (PDG), too. In comparison
to $\Lambda$CDM, the evolution of ultra-relativistic heavy-ion collisions is more complicated since different evolution stages (initial conditions, 
pre-equilibrium, hydrodynamization, hadronization, freeze-out, ...) are followed with different dynamical concepts, and since modeling is currently
needed to bridge between the different stages and between data and QCD-based predictions. The demonstration that such a more involved 
analysis can yield {\it PDG-grade} results could add further value to a central HIP effort, and it could motivate to scrutinize further in which
sense and with which precision the model parameters constrained in phenomenological analyses {\it are} the properties of hot QCD.

{\bf The clash of the default pictures of particle production in pp and AA collisions}

The scope of HIP goes well beyond hot QCD. It reaches from the study of systems that are so 
small that thermalization is questionable to the analysis of high-momentum transfer processes that, 
by definition, are far off equilibrium. One recent set of LHC discoveries that has attracted attention
from HEP and HIP alike concerns the question of how to understand soft multi-particle production in hadronic collisions.
The HEP and HIP approaches to this question are {\it maximally} different: 

In HIP, the standard assumption is that interactions between the many physical degrees of freedom produced in AA collisions are so 
efficient that an expanding fluid forms in the collision region. As a consequence of fluid-dynamic evolution, spatial asymmetries in the
initial nuclear overlap (the almond shape of finite impact parameter collisions and event-wise fluctuations) translate into  
pressure gradients that manifest themselves in measurable ``flow'' anisotropies $v_n$ of the momentum distributions. 
In this context, a preferred value $\eta/s \ll 1$ 
of the shear viscosity over entropy ratio~\cite{Bernhard:2019bmu,Devetak:2019lsk,Auvinen:2020mpc,Nijs:2020roc} 
indicates that interactions may be so efficient that the very notion of particle-like degrees of freedom within the matter becomes
questionable. This is theoretically conceivable since
we know of classes of strongly coupled quantum field theories (with gravity duals) whose non-abelian plasmas are free of 
quasi-particle-like excitations~\cite{CasalderreySolana:2011us}. The default picture of AA is that of an {\it almost perfect fluid} with 
negligible mean free path and {\it close-to-minimal dissipative} properties. 

In marked constrast, in HEP, the standard assumption is that interactions between physical degrees of freedom produced in 
pp collisions are negligible. As sketched in Fig~\ref{fig1}a), partons are evolved perturbatively based on the kinematics of the
parent partons that split but without accounting for the partonic phase space density within which these splittings occur.
Also hadronic modeling is free of density-dependent effects. The physical degrees of freedom in such models may be
characterized as  {\it free-streaming while fragmenting}. The picture shares similarities with an {\it ideal gas} of infinite
mean free path and, \emph{a fortiori}, of maximal dissipative properties. 
Bjorken was the first to speculate that this becomes questionable when partonic phase space densities 
increase with center of mass energy  $\sqrt{s_{\rm pp}}$. As 
early as 1982, he suggested that the outgoing parton showers could undergo secondary interactions with the underlying event
which could result in extreme dijet imbalances that he referred to as ``jet extinction''~\cite{Bjorken:1982tu}. 

HEP's default picture is at the basis of multi-purpose event generators that account for many characteristics of soft multi-particle
production in pp  collisions~\cite{Skands:2014pea,Field:2005sa}. But to check that it does not account for AA collisions, 
 it is sufficient to walk to the counting house of an LHC experiment and to stare at 
the event display (cf. Fig~\ref{fig1}b).  In the default picture of Fig.~\ref{fig1}a, multiple partonic interactions (MPIs)
contribute incoherently to the event multiplicity. 
In AA collisions, however,
azimuthal asymmetries $v_n$ in the distribution of $O(10'000)$ particles are visible by eye and they
cannot possibly result from an incoherent superposition of the $O(1000)$ nucleon-nucleon collisions that build up a central Pb+Pb event.
The abundance and robustness of such flow signals in heavy-ion collisions makes them one of the most important tools for studying 
collective phenomena in the quark gluon plasma~\cite{Shen:2020gef,Gajdosova:2020nvb}.
This was known for lower collider (RHIC) and fixed target (CERN SPS, BNL AGS) energies. Combined with the phenomenological
success of models that include fluid-dynamic expansion~\cite{Bernhard:2019bmu,Devetak:2019lsk,Auvinen:2020mpc,Nijs:2020roc}, it forms 
the main support for an almost perfect fluid picture of AA collisions.  Since the azimuthal asymmetry in Fig~\ref{fig1}b provides direct evidence 
for final state interactions,  Bjorken's reasoning is unavoidable. Indeed, strongly enhanced dijet imbalances ~\cite{Aad:2010bu,Chatrchyan:2011sx} and hadronic back-to-back correlations \cite{Adler:2002tq} have been observed in heavy ion collisions.

The dichotomy between the traditional pictures of particle production in pp and AA collisions has been called into question recently
by several LHC discoveries of heavy-ion like behavior in small  pPb and 
high-multiplicity pp collisions. In particular, in a sense that can be made statistically precise within the so-called cumulant analysis, 
LHC experiments discovered that the smaller azimuthal anisotropies observed in pp and pPb collisions are collective 
phenomena~\cite{Abelev:2014mda,Aaboud:2017acw,Sirunyan:2017uyl} that are inconsistent with an incoherent superposition of
MPIs.  Also, the assumption underlying  Fig.~\ref{fig1}a that 
hadronization is process-independent and thus independent of event multiplicity contradicts data from pp, pPb and PbPb collisions~\cite{ALICE:2017jyt}. None of the existing multi-purpose event generators for pp collisions can account for these classes
of phenomena, and 
one is forced to conclude that {\it ``more physics mechanisms are at play in proton-proton collisions than traditionally thought''}~\cite{Fischer:2016zzs}. Different additional mechanisms are currently explored~\cite{Bierlich:2018xfw,Fischer:2016zzs};
their common denominator is a non-trivial dependence on the density of physical degrees of freedom in pp collisions. 

{\it Does the LHC discovery of collectivity in small (pPb and high-multiplicity pp) collisions imply that a close-to-perfect fluid with QGP-like properties is
formed in these smaller systems, too? Or can systems of fundamental quantum fields exhibit significant flow-like signatures even if they
are too small and short-lived to hydrodynamize? } 
Good qualitative arguments support either viewpoint. On the one hand,  the many flow-like signals measured in pp and pPb collisions 
(harmonic coefficients $v_n\lbrace{2\rbrace}$, $v_n\lbrace{4\rbrace}$, $v_n\lbrace{6\rbrace}$, their $p_\perp$-dependence and mutliplicity
dependence, etc.) show the main signatures expected for fluid dynamic evolution with minimal dissipation. For pPb collisions, this 
dependence had been predicted in hydrodynamic simulations~\cite{Bozek:2011if} prior to any data taking. On the other hand, it remains unclear
how  to extend the applicability of hydrodynamics from large and dense systems, where interaction rates may be expected to maintain 
local equilibrium, to small, more dilute and short-lived systems, where this assumption seems to be more and more difficult to maintain~\cite{Schukraft:2017nbn}. 
These open conceptual issues have motivated many studies of how strongly expanding, far-out-of equilibrium systems
of fundamental quantum fields approach hydrodynamic behavior (i.e., ``hydrodynamize'' and thermalize locally). Here, one important novel concept
are hydrodynamic attractor solutions that can be approached by expanding systems well before the conditions of hydrodynamization 
and thermalization are met (see~\cite{Berges:2020fwq} and references therein). 

For a phenomenology across system size, the central question is: {\it How does strong interaction physics transit from an ideal gas to an almost 
prefect fluid as a function of the size of the collision system, its event multiplicity or other experimentally accessible parameters?} This has renewed 
interest in kinetic transport which is a candidate framework for reconciling the maximally different traditional pictures of multi-particle production in pp and AA, 
as it interpolates between free-streaming in very small and dilute systems and fluid dynamic behavior in sufficiently large and dense 
systems. In kinetic theories, infrequent rescatterings can result in large flow-like signals~\cite{He:2015hfa}.  Even in the case of only 
one rescattering, it is possible to account not only for the relation between initial spatial eccentricities $\epsilon_n$ and final momentum 
anisotropies $v_n$, but also for the non-linear mode-mode couplings that had been regarded previously as a hallmark of hydrodynamics~\cite{Borghini:2018xum,Kurkela:2018ygx}. Phenomenological partonic~\cite{Xu:2011jm,Xu:2004mz} and hadronic~\cite{Bleicher:1999xi,Petersen:2018jag,Sambataro:2020pge} transport models continue to be explored in 
comparison with LHC data; in parallel, there is progress on weakly coupled QCD kinetic theory (see~\cite{Berges:2020fwq,Du:2020dvp}
and references therein) and Boltzmann transport equations are studied in isolation to gain insights into how the transition
from free-streaming to perfect fluidity occurs. These studies make it now conceivable that significant flow-like
signals in small systems have a non-fluid dynamic origin.

\begin{figure}[t]
  \centering
\includegraphics[width=0.41\textwidth]{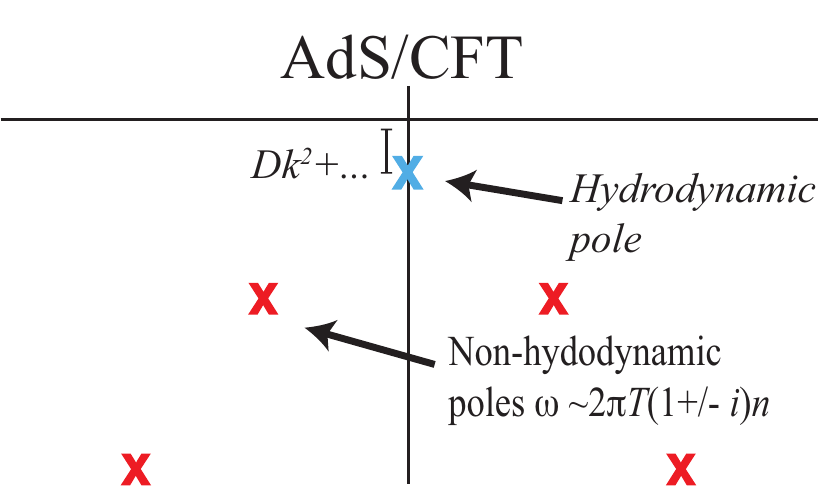}
\includegraphics[width=0.41\textwidth]{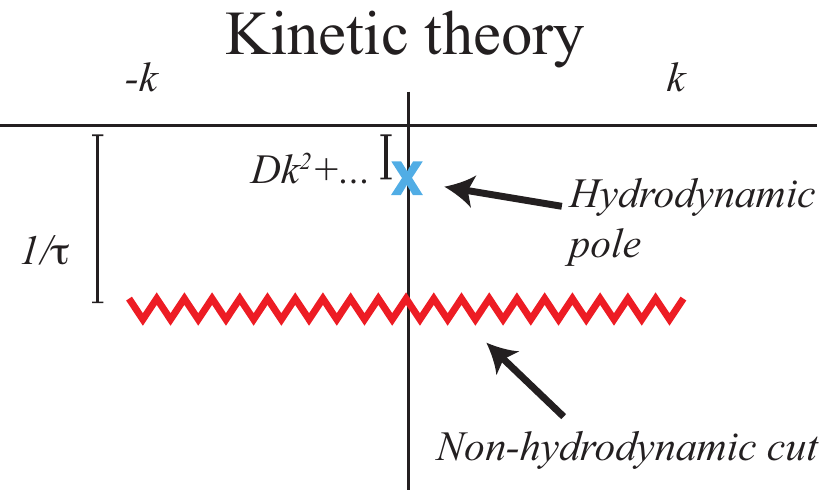}\\
\vspace{0.01\textheight}
  \centering
\includegraphics[width=0.41\textwidth]{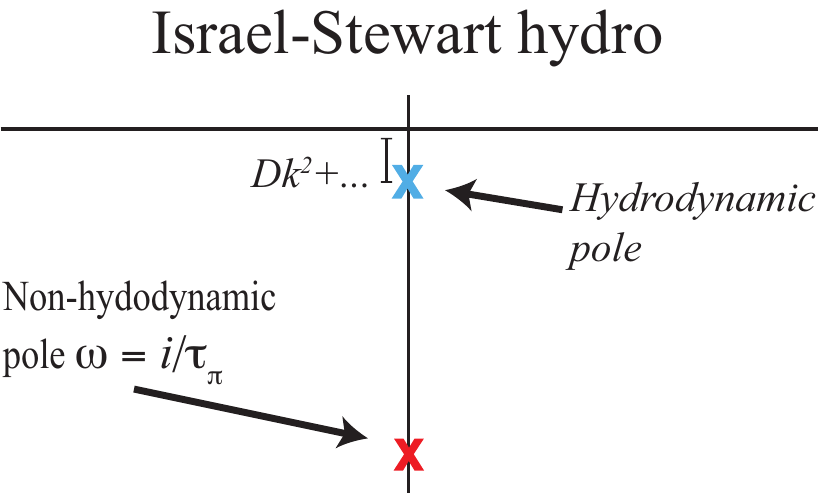}
\includegraphics[width=0.41\textwidth]{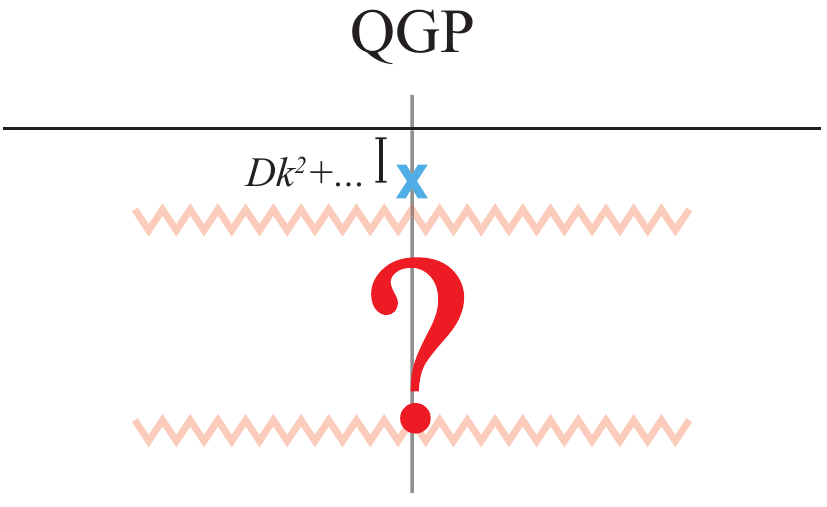}
\caption{Four sketches of the non-analytic
structures (poles and branch-cuts) of the retarded propagator of the energy momentum tensor in different plasmas. 
These structures are  in one-to-one correspondence with the physical degrees of freedom operational in the plasma. 
 Hydrodynamic excitations (blue) are a common feature of all Lorentz symmetric theories 
with self-interactions. Non-hydrodynamic excitations depend on the nature of the plasma and could include quasi-normal modes
(AdS/CFT), non-propagating dissipative excitations (Israel-Stewart hydro) or particle-like excitations represented by a
branch cut (kinetic theory). Discriminating between these possibilities is tantamount to establishing the microscopic
structure of the plasma. 
Figure taken from~\cite{Kurkela:2018ygx}.
}
\label{fig2}
\end{figure}

{\bf Why does it matter?}

The current discussion of how to understand best the LHC discovery of heavy-ion like behavior in small pp and pPb systems is mainly pursued
by comparing model scenarios that incorporate different physics assumptions. Model studies are clearly important, but they are technically
involved and difficult to communicate to a general audience. I therefore restrict myself here to simplified parametric considerations that 
may illustrate some of the fundamental theoretical concepts that are at stake. 

For this purpose, let us view a pp, pPb or PbPb collision as producing a system in the collision region that is described by a local 
energy-momentum tensor $T^{\mu\nu}(x,t)$. Experiments have learnt how to characterize the impact parameter and event-by-event
fluctuations of hadronic collisions, and we therefore assume that we have some way of introducing perturbations $\delta T^{\mu\nu}$ 
on top of the average $\langle T^{\mu\nu} \rangle$. We want to understand the material properties of the produced plasma 
by following the propagation of these perturbations $\delta T^{\mu\nu}$. Theory tells us that this propagation is defined by the retarded 
propagator $G_R^{\mu\nu,\alpha\beta}(x,t)$ of $T^{\mu\nu}$ and that there is a one-to-one correspondence between the physical degrees of 
freedom operational in the plasma and the non-analytic structures of the Fourier transformed propagator $\tilde{G}_R^{\mu\nu,\alpha\beta}(k,\omega)$ 
in the complex $\omega$-plane. Different physics pictures of the plasma correspond to different non-analytic structures, see~Fig.~\ref{fig2}.

For instance, a shear viscous hydrodynamic excitation corresponds to a pole at $\rm{Im}(\omega) = - \Gamma_s \, k^2$. Here, 
$\Gamma_s = \tfrac{\eta}{sT}$ is the sound attenuation length that determines how a perturbation $\delta T^{\mu\nu}$  attenuates in time  
 $\propto \exp\left[ - i\, \omega_{\rm pole} t \right] = \exp\left[ - \Gamma_s \, k^2\, t \right]$. One sees that a plasma of minimal $\eta/s$ is
maximally transparent to the propagation of $\delta T^{\mu\nu}$. Indeed, for a given initial spatial eccentricity $\delta T^{\mu\nu}$, 
the maximal final momentum anisotropy $v_n$ results for a minimal sound attenuation length. This may illustrate why the large
experimental values for $v_n$ imply $\eta/s \ll 1$~\cite{Bernhard:2019bmu,Devetak:2019lsk,Auvinen:2020mpc,Nijs:2020roc}. 

For reasons of causality, all forms of matter show also non-hydro excitations; fluid dynamics without non-hydro excitations 
is not realized in quantum field theory. Non-hydro modes are shorter-lived and thus more difficult to access experimentally, but they are of particular interest since they differ characteristically for different forms of matter (cf. Fig.~\ref{fig2}).
Consider e.g. a system of size $R$.
The smallest wave-number supported by this system is $k \sim 1/R$, and the typical time for in-medium propagation of any excitation is 
 $t \sim R$. For an excitation that propagates either hydrodynamically or non-hydrodynamically with poles at
 $\rm{Im}(\omega) = - \Gamma_s\, k^2$ and $\rm{Im}(\omega) = -1/\tau_R$, respectively,
 \[
 G_{R}(t, k) \propto \underbrace{ \left(  c_{\mathrm{hyd}} \exp \left[-\Gamma_s k^{2} t\right] \right.}_{\text {reduced for smaller } \mathrm{R}}+\underbrace{\left.c_{\mathrm{non}-\mathrm{hyd}} \exp \left[-t / \tau_{R}\right] \right)}_{\text {enhanced for smaller } \mathrm{R}}\, .
 \]
 For decreasing system size $R$, the exponential suppression ($\Gamma_s k^2 t \sim \Gamma_s/R$) of the longest wavelength
 hydrodynamic excitation {\it increases} while the one of the non-hydrodynamic excitation ($t/\tau_R \sim R/\tau$) {\it decreases}.  
 Varying the system size $R$ is thus a tool 
 that enables us to change the relative importance of hydro- and non-hydro excitations.  This parametric argument is
 supported by model studies in which one leaves the fluid dynamic sector (i.e. the equation of state and the 
 dissipative properties that fix the hydrodynamic poles in Fig.~\ref{fig2}) unchanged while replacing the non-hydrodynamic 
 pole of Israel-Stewart viscous fluid dynamics with the corresponding quasi-particle structure of a  kinetic theory. 
 One then finds that two different dynamical models with identical initial conditions, identical
 hydrodynamics and even identical relaxation times, can give rise to quantitatively different flow $v_n$ depending on 
 the physical nature of  their shortlived non-hydrodynamic excitations;
 these differences increase with decreasing system size  (more precisely: with decreasing opacity)~\cite{Kurkela:2018ygx}. 
 This indicates that experiments are sensitive to the nature of the non-hydro excitations in the QCD plasma.

{\bf Far-off equilibrium is hard}

 Any {\it hard} (\emph{i.e.} high-momentum transfer) process is initially
 far from equilibrium. In very small and dilute collision systems, hard partons shed off their initially high 
 virtuality in a {\it vacuum} parton shower whose hadronic fragment distributions can be characterized by jet measurements. 
 But imagine that the same hard parton is injected into a thermalized QGP. This parton will fragment, too, but it cannot hadronize. 
 Rather, after traveling some distance, the partonic
 fragments of this hard perturbation will be indistinguishable from the thermal constituents of the QGP: the parton has thermalized. 
 In this sense, Bjorken's jet extinction is jet thermalization, and the experimentally observed jet quenching is the characterization of a  thermalization process that was stopped prior to completion at a time-scale of the jet's in-medium pathlength. 
 Qualitatively,  thermalization of a hard (short wavelength $1/k$) perturbation proceeds by 
 isotropization and softening. Both are ubiquitous features of jet quenching measurements that show medium-induced enhancements 
 of soft fragments at large (isotropized) angle far outside typical jet opening cones. 
 Conceptually, the same microscopic interactions invoked in phenomenologically validated jet quenching models
 (the non-abelian Landau-Pomeranchuk- Migdal $1\to 2$ process and elastic $2\to 2$ scattering), are known to give rise to a perturbative thermalization and hydrodynamization mechanism if applied to oversaturated partonic systems~\cite{Baier:2000sb,Kurkela:2015qoa}.

\begin{figure}[t]
  \centering
\includegraphics[width=0.55\textwidth]{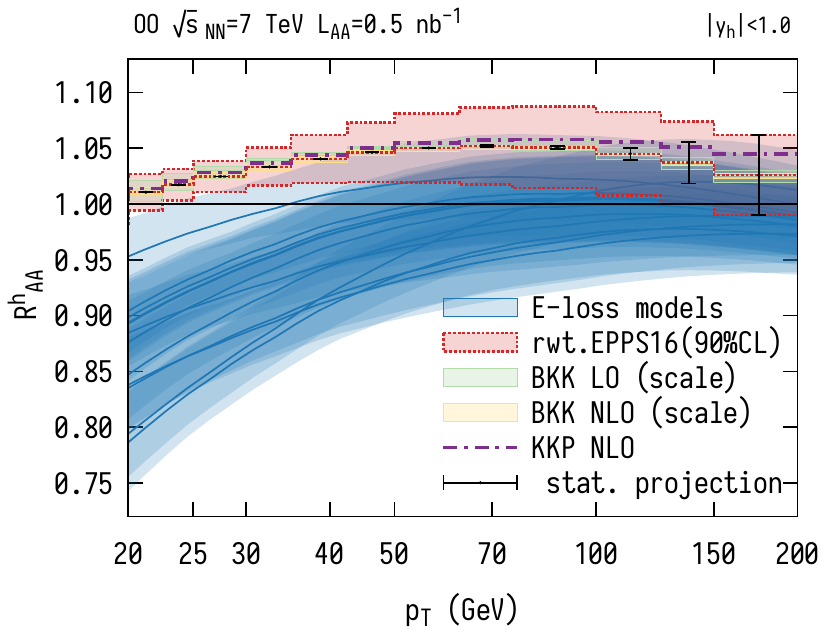}
\caption{For the hadronic nuclear modification factor
$R_{\mathrm{AA}}^{h}\left(p_{\perp}, y\right)=\frac{1}{A^{2}} \frac{d \sigma_{\mathrm{AA}}^{h} / d y d p_{\perp}^{2}}{d \sigma_{p p}^{h} / d y d p_{\perp}^{2}}$ in inclusive oxygen-oxygen collisions, systematic uncertainties in the no-jet quenching baseline (red band) are very small 
and can be fully quantified in NLO perturbative QCD. For small collision systems where jet quenching has escaped unambiguous detection
so far, this nuclear modification factor provides currently the highest sensitivity for future searches.
Essentially all jet quenching models (blue bands) can be separated from the null-hypothesis in  a short OO run.
Figure taken from~\cite{Huss:2020whe}.
}
\label{fig3}
\end{figure}

 From what we know, jet quenching is a precursor of thermalization and hydrodynamization. If true, this implies that it must occur 
 whenever flow-like phenomena are observed. For sufficiently central nucleus-nucleus collisions, the concurrence of jet quenching 
 and flow is well-established. But jet quenching becomes small in small systems, and uncertainties in its
 characterization increase with decreasing system size. As a consequence, jet quenching has escaped an unambiguous detection
 in small collision systems, including pPb collisions~\cite{ALICE:2012mj} where flow-like phenomena 
 are large. Why this is so is currently one of the most important open questions in the field. It may be addressed by 
 the standard HEP search strategy for the discovery of small effects, namely to formulate as precisely 
 as possible a systematically controlled null-hypothesis baseline that depends only on NLO pQCD, and to search then for jet quenching signals on top of it~\cite{Huss:2020whe}. Inclusive oxygen-oxygen collisions  seem particularly suited to this end
 since they are free of soft physics assumptions associated with the centrality selection in peripheral heavy ion collisions. As seen in
 Fig.~\ref{fig3}, the accuracy of this baseline can be known up to a few percent.
 It should then be possible to discriminate the baseline from essentially all jet quenching models that are currently considered.
 
 There are many other examples of how HEP techniques are currently retailored to address central questions in HIP. For instance, 
 HIP has started to employ modern jet finding algorithms~\cite{Andrews:2018jcm} to identify medium-modified parton splittings from 
 jet substructure. While HEP and HIP use essentially the same techniques for characterizing parton showers, they address very different 
 challenges: In HEP, the spatio-temporal ordering of the vacuum parton shower cannot be constrained experimentally, and it is generally
 regarded as being ambiguous. This is so, since a probabilistic interpretation of the vacuum parton shower is technically useful but 
 not unique - the only physical requirement is that the vacuum shower is parametrically correct to the desired logarithmic accuracy. 
 The main focus is then on increasing this accuracy. 
 In HIP, on the contrary, the spatio-temporal embedding of the parton shower in the medium affects the observable outcome: whether
 a  $1 \to 2$ splitting occurs sufficiently early determines whether one or two partons interact with the medium. Jet substructure analyses 
 constrain then not only {\it whether} a $1 \to 2$ splitting occured, but also {\it where} and {\it when} it occured within
 the collision region. (This statement captures the essential, but a proper theoretical formulation is more subtle since $1 \to 2$ is 
 a quantum mechanical decoherence process that extends over a finite region in space and time.) HIP thus adds many conceptually
 new facets to jet studies: i) it uses the quenching of jets to learn about the properties of the medium~\cite{Burke:2013yra}. ii) it uses the 
 medium to learn about the spatio-temporal structure of the jet (see e.g.~\cite{Apolinario:2020uvt}) and iii) it uses jet shape measurements 
 in medium to characterize isotropization  and
 softening which are hallmarks of the onset of thermalization processes for far-off equilibrium probes.

{\bf New Physics (NP)}

Irrespective of whether it lies within or beyond the current standard model, I would argue to use the notion {\it ``new physics at the LHC''}
for all physics that we were unable to test prior to the start of the LHC. Here, I have focussed on NP topics in HIP 
that relate to fundamental questions about the thermal sector of the standard model. In the second decade of the common journey
of HEP and HIP at the LHC, these and other topics will be advanced by an order-of-magnitude increase in the luminosity of PbPb
collisions and by exploiting the complementary information accessible in the collision of lighter ion beams and proton-nucleus collisions
(see Ref.~\cite{Citron:2018lsq} for documentation). I am excited about the increasing
number of topics on which both, HEP and HIP, will put their ink together.

I thank Federico Antinori, Jasmine Brewer, Aleksi Kurkela, Aleksas Mazeliauskas and Wilke van der Schee for discussions.

\end{document}